# Characterization of cesium and H$^-$/D$^-$ density in the negative ion source SPIDER

Marco Barbisan[a], R. Agnello[a,b], L. Baldini[c], G. Casati[d], M. Fadone[a], R. Pasqualotto[a,e], A. Rizzolo[a], E. Sartori[a,c] and G. Serianni[a,e]

[a] *Consorzio RFX (CNR, ENEA, INFN, Università di Padova, Acciaierie Venete SpA), C.so Stati Uniti 4, 35127 Padova, Italy*

[b] *École Polytechnique Fédérale de Lausanne (EPFL), Swiss Plasma Center (SPC), CH-1015 Lausanne, Switzerland*

[c] *Università degli Studi di Padova, Via 8 Febbraio, 2 - 35122 Padova, Italy*

[d] *Imperial College London, Exhibition Rd., South Kensington, SW7 2BX, London UK*

[e] *Istituto per la Scienza e la Tecnologia dei Plasmi, CNR, Padova, Italy*

E-mail: marco.barbisan@igi.cnr.it

ABSTRACT: The Heating Neutral Beam Injectors (HNBs) for ITER will have to deliver 16.7 MW beams of H/D particles at 1 MeV energy. The beams will be produced from H$^-$/D$^-$ ions, generated by a radiofrequency plasma source coupled to an ion acceleration system. A prototype of the ITER HNB ion source is being tested in the SPIDER experiment, part of the ITER Neutral Beam Test Facility at Consorzio RFX. Reaching the design targets for beam current density and fraction of coextracted electrons is only possible by evaporating cesium in the source, in particular on the plasma facing grid (PG) of the acceleration system. In this way the work function of the surfaces decreases, significantly increasing the amount of surface reactions that convert neutrals and positive ions into H$^-$/D$^-$. It is then of paramount importance to monitor the density of negative ions and the density of Cs in the proximity of the PG. Monitoring the Cs spatial distribution along the PG is also essential to guarantee the uniformity of the beam current. In SPIDER, this is possible thanks to the Cavity Ringdown Spectroscopy (CRDS) and the Laser absorption Spectroscopy diagnostics (LAS), which provide line-integrated measurements of negative ion density and neutral, ground state Cs density, respectively. The paper discusses the CRDS and LAS measurements as a function of input power and of the magnetic and electric fields used to reduce the coextraction of electrons. Negative ion density data are in qualitative agreement with the analogous measurements in Cs-free conditions. In agreement with simulations, Cs density is peaked in the center of the source; a top/bottom non uniformity is also present. Several effects of plasma on Cs deposition and negative ion production are presented.

# 1. Introduction

ITER Heating Neutral Beam injectors (HNBs) have to provide 16.7 MW beams of H/D particles at about 1 MeV energy. The high energy neutrals will be produced by gas neutralization of $H^-/D^-$ ions, generated in a source based on a radiofrequency (RF) inductively coupled plasma (ICP) generation concept [[1]-[4]]. The negative ions are extracted from the plasma in the source and accelerated by means of a system of grids. The current density of ions extracted from the source must reach target values of 330 A/m$^2$ (H)/290 A/m$^2$ (D), from a total extraction area of 0.2 m$^2$, and the ratio between coextracted electrons and negative ions should be lower than 0.5 (H)/1 (D); beam extraction must be continuously kept up to 1 h. In order to reach these challenging targets, a full scale prototype of the ITER HNB ion sources was built and is in operation at the ITER Neutral Beam Test Facility (NBTF), located at Consorzio RFX (Padua, Italy).

In the plasma volume inside the source, negative ions are generated by dissociative electron attachment reactions, involving rovibrationally excited $H_2/D_2$ molecules. In order to reach the target on the extracted current density, the production of $H^-/D^-$ must rely also on surface reactions, which convert H/D and $H^+/D^+$ in negative ions. Evaporating Cs on the surfaces of the source, and then lowering their work function, the surface reaction rate can be exponentially increased, making it largely dominant over volume reactions [[5]-[8]]. In this manner, at the extraction regions the plasma quasi-neutrality is guaranteed by the large negative ion density that is locally produced by surface conversion; in this regard, the electron density is minimised also thanks to the largely reduced electron mobility caused by the presence of a magnetic filter field. As a result, the co-extracted electron current is minimised. A key target is then that the work function on the plasma facing grid of the acceleration system is kept as low and uniform as possible. This is not just determined by where and how much Cs is actively deposited, but also by the action of the plasma, which erodes Cs from the surfaces and redistributes it inside the source.

The negative ion density in SPIDER was firstly characterized in Cs-free conditions, during the experimental campaigns up to April 2021 [[4],[9]]; the results were presented in refs. [[10]-[12]]. During the subsequent experimental campaign, from May to July 2021, Cs was evaporated for the first time in SPIDER [[13],[14]].

Object of this study is the behaviour of both Cs evaporation and negative ion density, in relation to the source parameters and to the beam properties (ion current and fraction of coextracted electrons). This is possible thanks to the Cavity Ring-Down Spectroscopy (CRDS) [[15],[16]] and Laser Absorption Spectroscopy (LAS) diagnostics [[17]-[19]], which provide line-integrated measurements of $H^-/D^-$ density and neutral Cs density at ground state, respectively. These diagnostic techniques are used in several negative ion sources for fusion and related facilities [[20]-[32]]. Since SPIDER is the largest negative ion source of its kind, attention will be given to spatial non uniformities, within the measurement capabilities of the diagnostics. Section 2 of this paper will describe the structure of the SPIDER source and its functional principles. Section 3 will briefly recall the structure and the data analysis methods of the CRDS and LAS diagnostics in SPIDER. At last, sec. 4 will present and discuss the experimental results, from the characterization of Cs evaporation in the vacuum phase to the effects of the main source parameters during the plasma phase.

## 2. The SPIDER ion source

The lateral view of the SPIDER negative ion source is schematically shown in Figure 1a. The plasma is generated in eight ICP plasma sources (hereafter called plasma *drivers*), arranged in four rows; each one of these rows of drivers is powered by a RF generator (max. 200 kW at 1 MHz), labelled RF1-RF4 [33]. The plasma in the drivers diffuses in a common plasma chamber, in order to get spatially uniform; surface plasma losses are limited by a multicusp magnetic filter field at the side walls. On the opposite side of the drivers, the plasma reaches the Plasma Grid (PG), from which negative ions are extracted.

In the backplate of the expansion chamber, in between the rows of plasma drivers, three ovens are installed to evaporate Cs in the source [13]. In each oven, the Cs is contained in a reservoir, from which vapours can propagate, through a duct, towards the nozzle, which juts out of the backplate surface. The Cs flux from each oven is determined by the temperatures of reservoir (mainly) and duct, which are actively controlled and stabilized [[34]-[36]]; the Cs flux is measured on each nozzle by a Surface Ionization Detector (SID) [37]. To avoid direct evaporation towards the PG apertures and therefore into the accelerator, where caesium at the electrode surfaces would affect the voltage holding during beam operation, the apertures on each nozzle let the caesium atoms flow laterally. Therefore, transport of caesium to the PG in SPIDER relies in part on the equilibrium in vacuum, because the fractional sticking would let caesium arrive to the extraction region in a few reflections, and in part on the redistribution caused by plasma-wall interactions.

In the expansion region, close to the PG, a transverse magnetic filter field is produced by means of a current $I_{PG}$, flowing vertically in the PG; all the data considered in this study were acquired with $I_{PG}$ in its standard direction, i.e. from top to bottom. The filter field intensity is 1.6 mT close to the PG per 1 kA of current, up to 8 mT [38]. Thanks to this magnetic field, the electrons diffusing from the drivers become magnetized; as result, the electron density and temperature get lower, from about $10^{18}$ m$^{-3}$ and 10 eV in the drivers to few $10^{17}$ m$^{-3}$ and 2 eV or less in proximity of the PG, respectively [[6],[39]-[43]]. The reduction of the electron temperature is essential to limit negative ion losses due to electron stripping [44]. The reduction of both electron density and temperature is also beneficial in limiting the coextraction of electrons from the PG apertures [45].

In order to lower the fraction of coextracted electrons as required in ITER, the electron density at the extraction apertures is further reduced by biasing the PG positively against the source; in this way, the PG attracts and removes more electrons, besides minimising the electric field that would otherwise push negative ions away from the extraction region [[6],[20],[45]-[50]]. Similarly, the source is equipped with a Bias Plate, an electrode placed at 10 mm from the PG and framing the groups of PG apertures. The BP is represented in Figure 1b in light blue; as with the PG, the BP is biased positively against the source body. The biasing of PG and BP is independently performed by the ISBI and ISBP power supplies, which are current-controlled in order to manage the net flux of removed electric charges. Both the power supplies have a resistor R=0.63 Ω in parallel, to handle both the current settings and the voltage imposed to PG and BP by the plasma; for low currents, PG and BP could be at a lower potential than the source body.

Negative ions, which are mostly produced at the upstream surface of the PG, can be extracted through its 1280 apertures. However, during the considered experimental campaign, only 28 apertures were active (red dots in Figure 1b), while the others were covered by a molybdenum mask. This was done to limit the gas outflow from the source, and so the gas pressure around the source and the consequent risk of breakdowns on the RF components [[4],[13],[14]]. In the active apertures, the negative ions are attracted by the extraction potential difference $U_{ex}$ between the second grid, the Extraction Grid (EG), and the PG. A set of magnets in the EG generates a magnetic filter field which dumps most of the electrons on the EG surface [[51]-[52]]. After having passed the EG apertures, the negative ions are accelerated up to the desired energy by the acceleration potential difference $U_{acc}$ between the final grid (Acceleration Grid – AG) and the EG. Thanks to this multiple gap approach it is possible to limit the energy of the coextracted electrons. $U_{ex}$ and $U_{acc}$ are set by the ISEG (Ion Source – Extraction Grid) power supply (maximum ISEG ratings: 12 kV, 140 A [33]) and by the AGPS power supply (Acceleration Grid Power Supply, maximum ratings: 96 kV, 71 A [53]). The maximum possible energy of negative ions in SPIDER is then 108 keV. In first approximation, the coextracted electron current $I_e$ can be estimated as the EG current, i.e. ISEG current minus AGPS current. The beam current $I_b$ exiting the AG was estimated from the instrumented calorimeter STRIKE, positioned at short distance (~0.5 m) from the GG [[13],[14]].

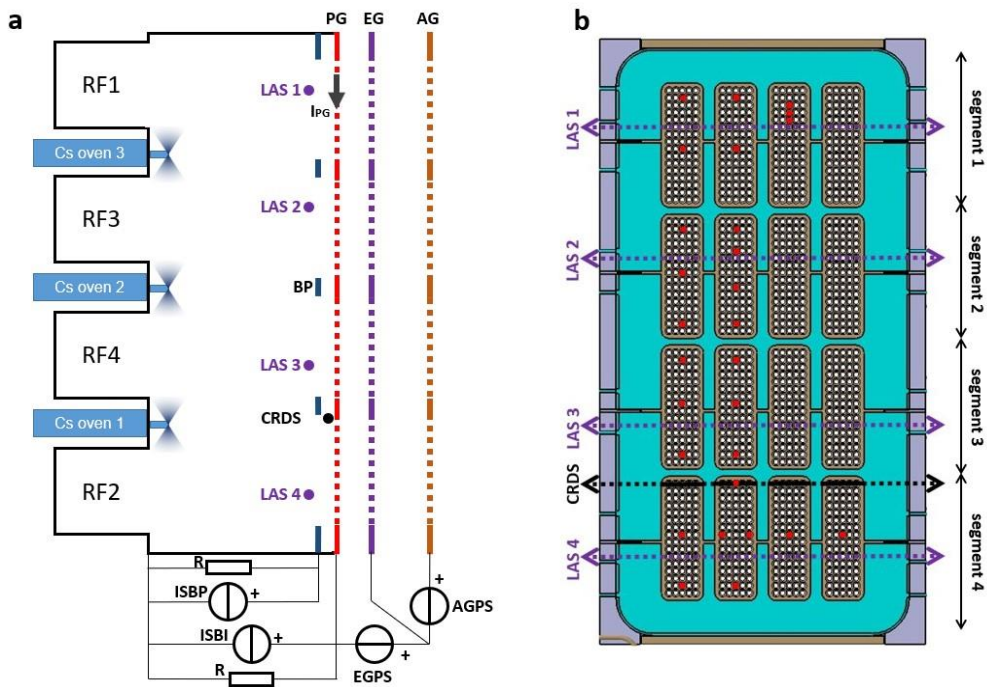

**Figure 1.** a) Scheme (vertical section) of the source and of the acceleration system of SPIDER, together with the electrical scheme of the main power supplies. b) Schematic representation of the BP (light blue) and of the PG, as seen from the back of the source; the active PG apertures are indicated with red dots, while the lines of sight of the LAS and CRDS diagnostics are indicated in purple and black, respectively.

## 3. Diagnostic setup

### 3.1 Cavity Ring-down Spectroscopy

The CRDS diagnostic in SPIDER [[10], [16]] is based on the principle that negative ions can absorb photons γ of sufficient energy (≥0.75 eV) to free the extra electron thanks to photo-detachment reactions (H$^-$/D$^-$+γ→H/D+e). In SPIDER, photons are provided by a Nd:YAG laser, emitting beam pulses at 1064 nm wavelength, 6 ns duration and 150 mJ energy. The photodetachment cross section ($\sigma=3.5\cdot10^{-21}$ m$^2$ [54]) would be too low to measure the light

– 3 –

absorption over a single pass through the negative ion volume in the source. Therefore, the laser pulse is trapped inside an optical cavity composed by two high reflectivity (>99.994%) mirrors, placed at opposite sides of the source. The laser pulse is then forced to travel back and forth in the plasma volume for thousands of times. The high reflectivity mirrors still let a minimal amount of light exit the source at each reflection. For each laser pulse entering the cavity a train of light pulses can be collected and measured in intensity on the other side. The envelope in time of each train of pulses shows an exponential decay, as shown in the example signal of Figure 2, collected by the detector during a vacuum phase; the signal polarity is reversed due to electronics, and the single pulses cannot be distinguished due to low-pass filtering. The decay time $\tau_0$ of the curve is estimated by a fitting function of the type $y = b - A \cdot \exp(-t/\tau_0)$, where b is the background level, A is the signal amplitude and t is time. As discussed in ref. [10], every experimental signal actually shows a multi-exponential decay from different cavity modes; fortunately, the spurious modes have much shorter decay time and rapidly disappear. For this reason, the first part of each ringdown signal is excluded from the fit procedure; the duration of the exclusion interval was empirically chosen to be 20 μs. With plasma and negative ions in the source, the photodetachment reactions cause a reduction of the decay time down to a level τ. A line-integrated estimate of negative ion density can then be calculated as

$$n_{H-} = \frac{L}{\sigma c d}\left(\frac{1}{\tau} - \frac{1}{\tau_0}\right) \quad (1)$$

where L=4.637 m is the distance between the high reflectivity mirrors, c is the speed of light and d=0.612 m is the estimate of the path length through the negative ion region. As explained in detail in ref. [10], d is much shorter than L due to the dimension differences between the source and the vacuum vessel, and because the line of sight is partly covered from the plasma due to the BP (see also Figure 1b). To compensate for drifts in the decay time, the $\tau_0$ values corresponding to the τ ones are estimated by linear fitting the available decay times in the vacuum phase over time [[10]-[12]].

In SPIDER one CRDS optical cavity/Line of Sight (LoS) is presently in operation. As shown in Figure 1, it is horizontally oriented in between PG and BP, at 5 mm from the upstream surface of the PG, over the highest row of apertures in the lowest PG segment(-429 mm in height from the PG center). More details on the diagnostic are available in refs. [[10]-[12]].

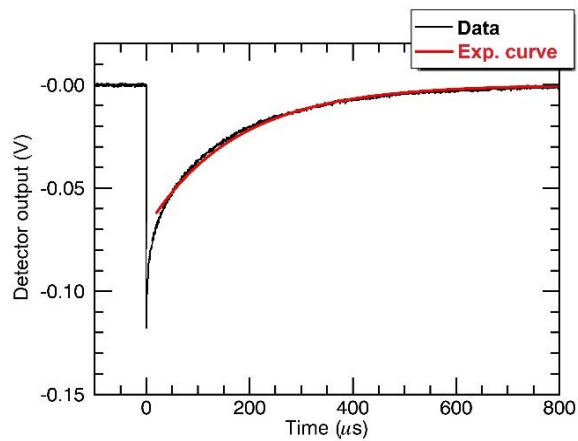

**Figure 2.** a) Example of CRDS signal acquired during a vacuum phase, as function of time. The red line is the fitted exponential curve.

## 3.2 Laser absorption Spectroscopy

In order to measure the density of neutral Cs (at ground state), the LAS diagnostic uses a Distributed FeedBack laser diode (DFB), whose emission wavelength can be finely tuned by changing supply current (3pm/mA) or diode temperature [[18],[19]]. While the temperature is kept stable by a PID system, the current is sawtooth modulated, so to make the laser scan a wavelength range around the Cs D2 line (852.1 nm, $6^2P_{3/2}$–$6^2S_{1/2}$ transition). Thanks to a system of fibers and collimators, the laser light is split into four optical paths, sent inside the source and collected on the other side, to be finally measured by as many detectors. The system was regulated so that less than 1 W/m$^2$ is emitted along each line of sight, preventing depopulation effects [[18],[19]]. As shown in Figure 1 a and b, the four horizontal lines of sight are at 5 mm and 25 mm distance from the upstream faces of BP and PG, respectively. The diameter of the laser beams is about 3.6 mm, while the diameter of the lines of sight set by the receiving collimators is about 9.5 mm. The lines of sight are equally distributed among the four PG segments (+649 mm, +253 mm, -253 mm and -649 mm from the height of the PG center) and labelled LAS1-LAS4 from top to bottom. Cs uniformity is essential for the uniformity of beam current; measuring the Cs density over the vertical direction was chosen because it's the same direction of the plasma drifts caused by magnetic and electric fields for electron suppression [[13],[14],[50],[55]-[60]].

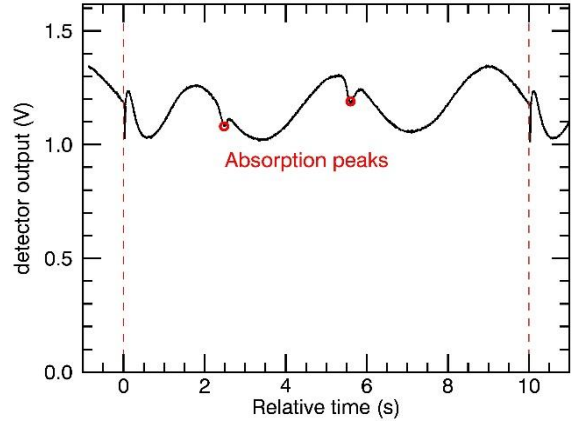

**Figure 3.** Example of LAS detector signal for a single wavelength scan, acquired during a vacuum phase. The red dashed vertical lines indicate the start and end of the scan.

An example of detector signal for a single wavelength scan is shown in Figure 3. The scan duration, which defines the temporal resolution of the diagnostic, was set to 10 s to allow the laser temperature PID to accommodate the thermal effects of the transition between the peak of a sawtooth current ramp and the immediately following current minimum. The absorption spectrum of the D2 line appears with its fine structure; two peaks are visible, each one including three transitions: F=3→F=2,3,4 and F=4→F=3,4,5. Knowing the separation of the peaks (21.4 pm), with a sawtooth modulation of the laser current and assuming a linear relationship between modulation signal and wavelength, the time base can be converted to wavelength[17]; the scan duration is sufficiently long to avoid nonlinearities that could arise from temperature oscillations, produced in turn by the laser current scanning. The Cs density at ground state $n_{Cs}$ can be calculated as follows:

$$n_{Cs} = \frac{8\pi c}{\lambda_0^4} \frac{g_k}{g_i} \frac{1}{A_{ik} l} \int ln\left[\frac{I(\lambda,0)}{I(\lambda,l)}\right] d\lambda \quad (2)$$

where $c$ is the speed of light, $\lambda_0 = 852.11$ nm is the D2 line wavelength, $g_k = 2$ and $g_i = 4$ are the statistical weights of lower and upper levels, $A_{ik} = 3.276 \cdot 10^7\ s^{-1}$ is the transition probability for the D2 line spontaneous emission, $l = 0.87\ m$ is the horizontal width of the source at the position of the lines of sight, $I(\lambda, 0)$ and $I(\lambda, l)$ are the intensity of the laser beam at wavelength $\lambda$ before and after having passed through the plasma; they are both derived from the detector signal, after careful subtraction of the signal background level in absence of light. $I(\lambda, l)$ is assumed to be the background-corrected detector signal itself, while $I(\lambda, 0)$ is estimated with a 9[th] degree polynomial fit of the corrected detector signal after having trimmed the absorption peaks. More specifically, under the typical laser modulation settings, ±4 % of the scan interval is ignored around each absorption peak; during plasma phases, the exclusion intervals are increased to ±8 % of the scan interval, because of the larger width of the absorption peaks. Once the coefficients of the polynomial fit are derived, they allow to obtain the $I(\lambda, 0)$ curve over the entire scan interval. In the $ln[I(\lambda, 0)/I(\lambda, l)]$ signal, the absorption peaks can be fitted with Gaussian curves, so that the spectral widths of the two peaks can be measured. The Doppler effect is known to cause a broadening of the spectral lines [61], given in this case by

$$\Delta w = k\lambda_0 \sqrt{\frac{T}{m_{Cs}}} \quad (3)$$

where $\Delta w$ is the line broadening (as wavelength) at Full Width at Half Maximum (FWHM), k=7.16·10[-7] K[-1/2] is a constant, T is the Cs temperature and $m_{Cs}$=132.91 is the Cs mass in atomic mass units. It is then possible to estimate the temperature $T_{Cs,p}$ of neutral, ground state Cs when the plasma is active:

$$T_{Cs,p} = \frac{m_{cs}}{k^2} \frac{w_{Cs,p}^2 - (w_{Cs,v}^2 - \Delta w_{room}^2)}{\lambda_0^2} \quad (4)$$

Where $w_{Cs,p}$ and $w_{Cs,v}$ are the average FWHM (wavelength) widths of the two peaks in the plasma and vacuum phase, respectively, while $\Delta w_{room}$ is the wavelength Doppler broadening calculated for room temperature (300 K). $(w_{Cs,v}^2 - \Delta w_{room}^2)$ allows to subtract from $w_{Cs,p}^2$ all the contributions to the line width that are not due to Doppler broadening (eg. intrinsic width, hyperfine structure, etc).

Further details on the LAS diagnostic in SPIDER can be found in refs. [[18],[19]].

## 4. Experimental results

For all the experimental measurements considered in the following subsection, the statistical measurement error is ±10 %.

### 4.1 Cs distribution

While the properties of the Cs layer on the PG cannot be directly characterized during the source operation, the density of Cs on its proximity can be sampled by the LAS diagnostic. Figure 4 shows Cs density values averaged over the four LoSs, and over intervals of time with fixed ovens evaporation settings and with stationary Cs density conditions. The measurements during the plasma phases, being significantly different from the vacuum phase, were not considered. The measured Cs densities are in any case the result of a dynamic equilibrium between vacuum phases and plasma phases, which cause a stronger Cs redistribution on the surfaces; for lines of sight that are very close to a surface (as in this case), the Cs density in the LoS volume is strongly affected from the Cs fluxes to and from that surface, and then indirectly on its Cs coverage. The Cs density values are shown as a function of the total Cs evaporation rate $r_{Cs}$ from the ovens, distinguishing conditions of uniform evaporation (the Cs rate of each oven differs from the others for less than 10%) in blue from slightly non-uniform evaporation conditions (at least two Cs rate values differ between each other for more than 10 %) in orange. In these experimental conditions the source body was kept at about 35°C, while the PG was set between 35°C and 140 °C, depending on the experimental session (the points in deuterium were taken with PG at 80°C). Besides the influence of Cs evaporation rate, the density was affected by the duty cycle of the plasma phases, the temperature of the PG, the action of impurities over time, etc. [[13],[62]] The plot shows that, given a total Cs evaporation rate, the <u>maximum</u> average Cs density is roughly linear (independently from evaporation uniformity at least below 24 mg/h in hydrogen). Such rough linearity may be due to the experimental mode of scaling the evaporation rate with the plasma duty cycle, aiming at stabilizing the extracted currents and implicitly the Cs conditioning of the PG (eg. doubling the plasma duty cycle is accompanied by a doubling of the Cs oven evaporation rate); this experimental technique is discussed and motivated in detail in ref. [13].To better understand the relationship between Cs density and evaporation rate, more measurements will be required in future, above 20 mg/h.

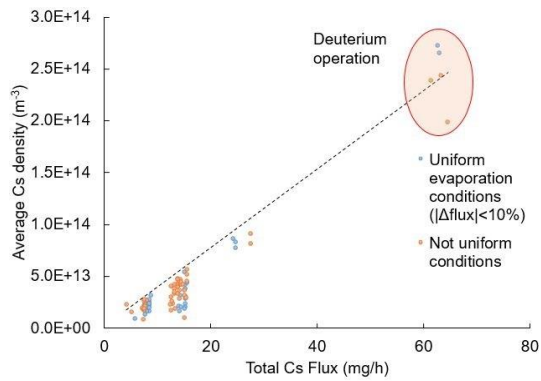

**Figure 4.** Average Cs density measured in vacuum phases by LAS as a function of total Cs evaporation rate from the Cs ovens. Blue and orange dots refer to uniform and non-uniform evaporation conditions (see text), respectively.

The uniformity of Cs evaporation is extremely important in determining the uniformity of beam current and, indirectly, beam divergence [63]. In order to find the best strategies for Cs evaporation, the experimentation on H$^-$ sources has always been accompanied by numerical simulations [[30],[64]-[66]]. The AVOCADO code allows to study the pressure distribution of a gas inside a vacuum environment. The validity of the code is ensured if the Knudsen number is greater than one: for Cs and assuming a maximum operating evaporating pressure of $10^{-4}$ Pa and a temperature of 100°C, the mean free path of the Cs particle is 41.2 m. Considering that plasma box height is less than 2 meters, the Knudsen number constraint is verified [67]. Figure 5a shows the Cs flux impinging on the PG, calculated assuming an equal Cs evaporation rate of 5 mg/h per each nozzle (15 mg/h in total). In the simulation, it is possible to assign a negative specific pumping speed to a surface in order to simulate a sticking coefficient, that is to simulate the probability of a Cs particle to bounce or to stay attached to a surface after hitting it. Thanks to the experience on the test stand in which the oven were initially tested [68], in this preliminary calculation a sticking coefficient of 2% was set on the source walls to predict SPIDER Cs oven behavior. Moreover, a different sticking value was set for the PG holes in order to simulate the incomplete transmission throughout the accelerator due to the presence of the remaining grids (14.5 %, assuming all the PG apertures open as first approximation; a sub reduced model has been studied for this purpose). Figure 5a shows also the effect of the BP masking on the Cs deposition on the PG. The results of AVOCADO could be used to calculate the line-integrated density of Cs, as observed from each LAS line of sight, keeping into account that the LoSs are upstream the BP and therefore not directly affected by the BP masking.

From the preliminary scenario presented in Figure 5a, as similarly done in [68], the sticking coefficient was progressively changed to match the typical Cs density measurements (roughly between $1 \cdot 10^{13}$ m$^{-3}$ and $5 \cdot 10^{13}$ m$^{-3}$) given by LAS at about 15 mg/h in total. For these progressive estimations, each tested sticking coefficient value has been set for all surfaces, except on the plasma grid apertures, where Cs atoms are assumed to definitely escape, i.e. a virtual 100 % sticking. Figure 5b shows the line-integrated density values calculated by AVOCADO for three sticking coefficient values (40%, 60 %, 80%) as a function of the vertical position y, and more specifically in correspondence of the four lines of sight (indicated with vertical lines). As shown by the figure, a sticking coefficient of 80% has to be set to roughly match the expected range of experimental data. This significant difference compared to the 2% found in the test stand can be explained by three factors: i) the presence of a plasma interacting with surfaces; ii) the conditions of SPIDER walls were not as clean as the ones of the test stand; iii) SPIDER plasma box walls are made of copper coated with Mo, while the test stand ones are made of stainless steel; iiii) SPIDER wall temperatures were variable throughout the experimentation and always higher than the test stand surface temperature (30°C). Finally, the AVOCADO results in Figure 5a show that Cs is expected to be predominantly deposited in the central region of the PG than in the more external areas, where the effect of the three ovens overlaps; this is reflected in the calculated Cs density values of Figure 5b, where LAS central LoSs are expected to measure higher densities than top and bottom LoSs. According to Figure 5a, a very localized reduction of Cs deposition would be expected in the exact center of the PG, likely because it coincides with the axis of the central Cs oven (the nozzle of Cs ovens emits laterally). However, this feature of Cs deposition is not expected to be measurable because of the vertical positions of the available LoSs.

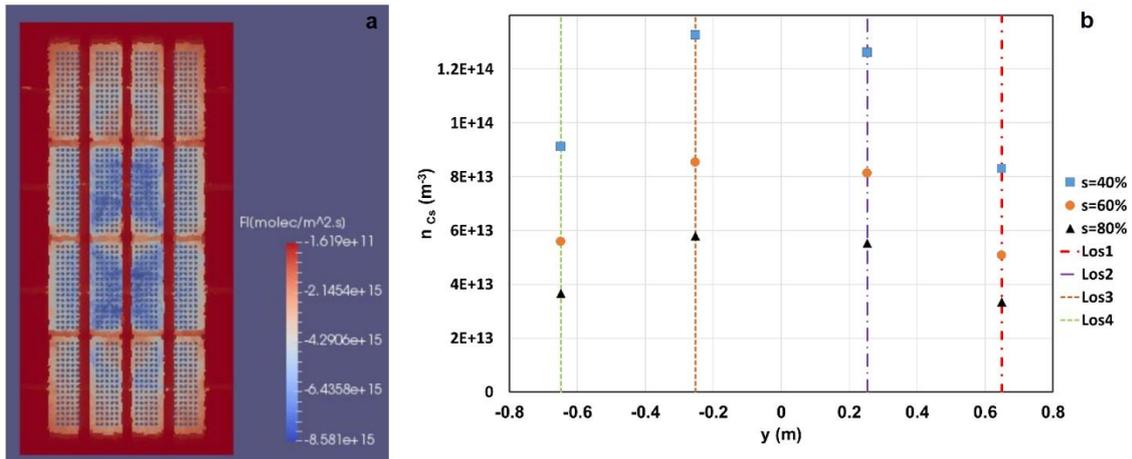

**Figure 5.** a) Cs flux on the upstream PG surface during a vacuum phase, as simulated by the AVOCADO code. The evaporation rate of each oven was set to 5 mg/h (15 mg/h total). The sticking coefficient of the source walls was set to 80%. b) line-integrated Cs density values calculated by AVOCADO for three sticking coefficient values (40%, 60 %, 80%) as a function of the vertical position y, in correspondence of the four LAS lines of sight (indicated with vertical lines).

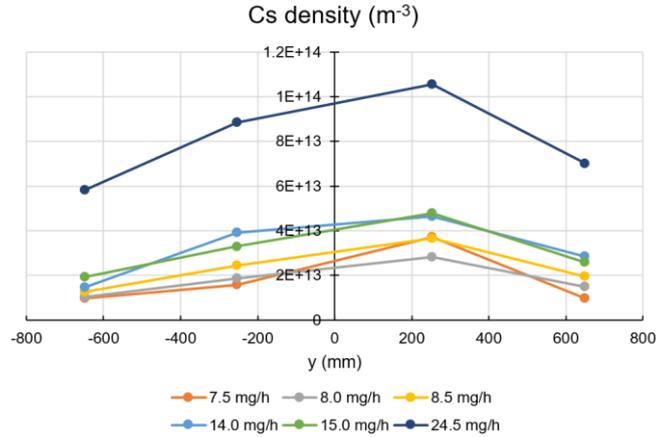

**Figure 6.** Experimental values of Cs density as a function of the LoS vertical position, for different values of evaporation rate, during vacuum phases in hydrogen operation Density values are averaged as described in the text.

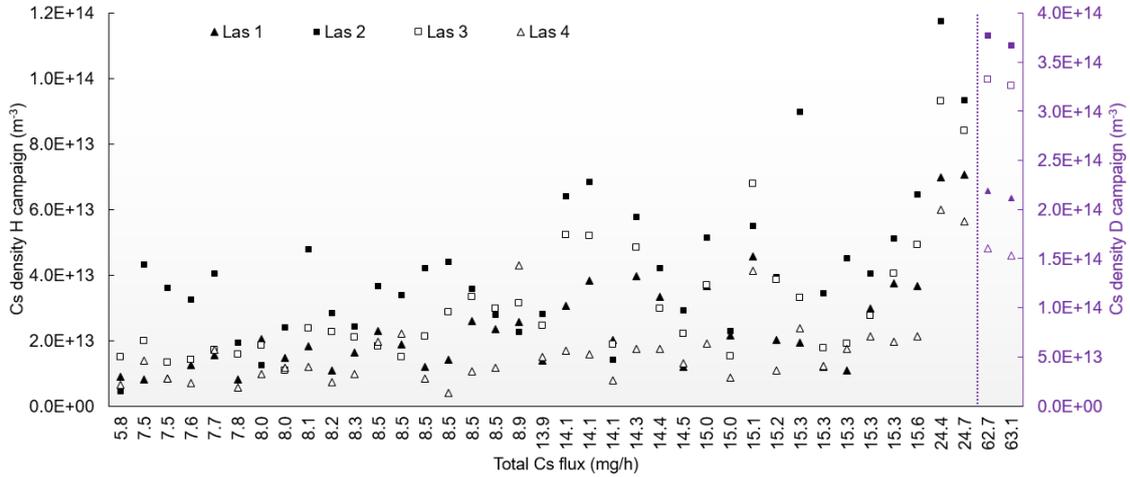

**Figure 7.** Cs density, as measured by each LAS LoS, under uniform evaporation conditions during vacuum phases; the cases are shown in ascending order with respect to the total evaporation rate. The values obtained in vacuum during the deuterium campaign, therefore with higher evaporation rates, are reported in a different scale in purple for visualization purposes.

The forecast of the simulations is basically conformed in Figure 6: the measured Cs densities are shown as a function of the LoS vertical positions. The shown data are averages of density measurements, over intervals of time with fixed and uniform evaporation rate among the three ovens (as previously defined), and under steady Cs density trend. In all these time intervals the evaporation rate did not differ by more than ±0.3 mg/h with respect value indicated in the legend. With respect to the simulations, an asymmetry is observed between the top and the bottom halves of the source: LAS1 is mostly higher than LAS 4 and LAS 2 is mostly higher than LAS 3. As consequence, LAS 4 has almost always the lowest values of Cs density. The top/bottom asymmetry may be due to the fact that the Cs oven 1 (bottom) always showed slightly lower values of evaporation rate (still with the 10 % limit imposed on the shown data). It is important to remind that the Cs layer builds up in vacuum for a longer time, but it is modified with stronger effect during the much shorter plasma phases. Therefore, the vertical distribution measured in vacuum may add to the action of the plasma, which is subject to vertical drifts. This was not simulated. More details and improved numerical studies are under way [69].

Figure 7 allows to get a better insight on each separate time interval with fixed and uniform evaporation rate, under steady Cs density trend, excluding plasma phases. The average values of Cs density for each separate LoS are reported for increasing values of $r_{Cs}$, indicated on the horizontal axis. Figure 7 also shows few cases in deuterium, reported in purple on a different scale for visualization purposes (deuterium operation require higher evaporation rates). The figure indicates that, even for similar evaporation rates (within reasonable errors on its estimation), the resulting caesium density is subject to non-negligible variations, due to the "history" of Cs evaporation. As previously discussed, the present status of the source surfaces is determined by the effect of cumulated caesium layer, plasma interaction and redistribution, quality of the background atmosphere and surface temperatures. Still, the overall features of the vertical Cs profile shown in Figure 6 are observed in most of the specific cases of Figure 7.

### 4.2 Influence of RF power

A basic parameter to control the negative ion production is the RF power, which has a major role in determining the plasma properties; the rate of $H^-/D^-$ ions generated at the PG depends on the flux of particles available for conversion on the surface. The plasma is also known to cause erosion processes that redistribute Cs on the surfaces of the source (PG included) [[13],[70]]. Moreover, the plasma ionizes a significant fraction of the Cs circulating in the source volume [17]; due to the electric field present in the volume, $Cs^+$ has a different transport than neutrals [30], attempts to estimate the Cs ionization degree are under way [71]. Previous campaigns in SPIDER have shown that, in Cs-free conditions, the negative ion density at the PG grows linearly with the RF power, as consequence of the parallel increase of plasma density on the volume reactions that generate negative ions [12]. Studies in other negative ion sources have verified this behaviour also with Cs evaporation [[21]-[23]]. Figure 8a shows the $H^-$ density (black dots) and the accelerated current density (orange dots, given by $I_b$ divided by the area of the active beamlets), as a function of the total RF power. The experimental conditions were: hydrogen gas at pressure $P_s$=0.4 Pa, magnetic filter field at the PG $B_{PG}$=2.4 mT, 80 A ISBI current, 0 A ISBP current, $U_{ex}$=5 kV, $U_{acc}$=40 kV, total evaporation rate $r_{Cs}$=7.5 mg/h; plasma pulses lasted 27 s, regularly repeated every 6 min.. The negative ion density grows with a basically linear trend with the RF power; the accelerated current density also follows a linear relationship with the RF power; considerations on the mutual relation between negative ion density and beam current density should take into account that the former is subject to spatial nonuniformities, due to plasma drifts and the masking effects of the BP [14], while the latter is an averaged measurement over the entire grids.

During the experimental campaign, the negative ion density was sensitive to the PG Cs conditioning; the maximum obtained negative ion density was $7.6 \cdot 10^{16}$ m$^{-3}$, reached with total RF power $P_{RF}$=400 kW, $P_s$=0.4 Pa, $B_{PG}$=1.6 mT, 80 A ISBI current, 80 A ISBP current and $r_{Cs}$=12 mg/h. Because of the above mentioned masking effects of the PG and of the position of the CRDS LoS (on the topmost row of apertures in PG segment 4, see Figure 1b) [14], slightly higher values of negative ion density may have been present in the central regions of the PG groups of apertures. The negative ion density values in SPIDER are in basic agreement with respect to other negative ion sources under Cs evaporation, keeping into account the different number of plasma drivers and the slightly different experimental conditions [[21]-[23]]. With respect to what found during the Cs-free campaigns in SPIDER, in presence of Cs the H⁻ density is about three times higher; a similar ratio was found with CRDS in the past in the BATMAN negative ion source [21].

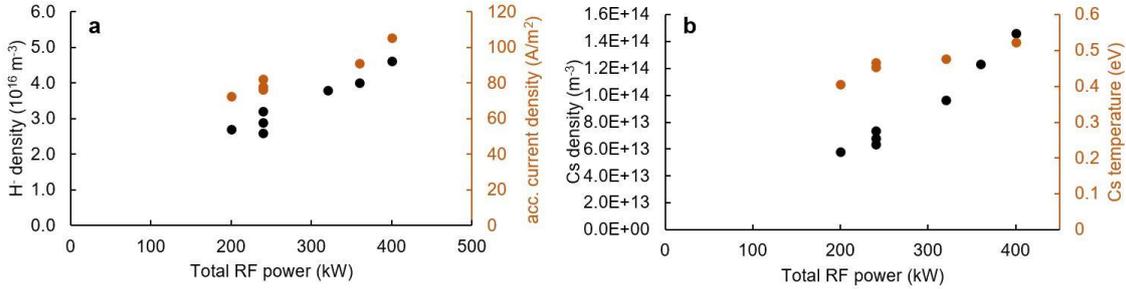

**Figure 8.** a) H⁻ density (black) and accelerated current density (orange) as a function of the total RF power. Experimental conditions: $P_s$=0.4 Pa (H$_2$), $B_{PG}$=2.4 mT, 80 A ISBI current, 0 A ISBP current, $U_{ex}$=5 kV, $U_{acc}$=40 kV, $r_{Cs}$=7.5 mg/h. b) LoS-averaged Cs density during the plasma phase and Cs temperature, as a function of total RF power. The plasma pulses are the same of plot b.

During plasma phases, the Cs density was always higher than in the vacuum phases that preceded or followed, indicating that, in terms of neutral Cs population, Cs erosion or reflection on the BP-PG surfaces dominates over the ionization reactions. For the same plasma pulses considered in Figure 8a, Figure 8b shows that the LoS-averaged Cs density increases as a function of RF power, analogously to what found in the ELISE test facility [66]. For higher and higher RF power ($P_{RF}$ should be increased in future up to 800 kW), the PG surface conditioning is expected to be more and more dependent on the action of plasma, especially for long plasma pulses (from the present 27 s for Cs conditioning up to 3600 s [1]). Besides Cs density, Figure 8b also shows the corresponding Cs temperature, measured from the Doppler broadening of LAS absorption spectra, indicating a slight increase with the RF power. The values, expressed in eV, correspond to about 4700 K÷6100 K, quite higher than the typical H$_2$ temperature measured during plasma pulses (around 1000 K [72]). A possible explanation is that part of the neutral Cs population is supplied by Cs⁺ ions, which gain energy from the electric fields in the plasma [73]. Cs⁺ particles can then neutralise on the source walls, including the surfaces close to the LoSs (PG and BP). A clear understanding of these measurements will require a complex modelling of the Cs distribution in presence of plasma in SPIDER.

### 4.3 Influence of magnetic filter field

In SPIDER, the combined use of the CRDS diagnostic and of RF compensated Langmuir probes on the BP showed that, in Cs-free conditions, the magnetic filter field can affect the plasma density and the electron temperature at the PG [12]. The negative ion density was maximized at about 1.4 mT÷1.6 mT magnetic filter field intensity, following the same behaviour of the plasma density at the BP. The effect of the plasma density peaking for a certain $B_{PG}$ was reinforced by the $B_{PG}$ influence on electron temperature: decreasing $B_{PG}$ below 1.2 mT, the electron temperature increased up to about 4.7 eV, significantly increasing negative ion losses.

During the campaign with Cs evaporation, it was only possible to keep the plasma stable with $B_{PG}$ in a limited range and for few cases. Figure 9 shows the negative ion density as a function of the magnetic filter field intensity, in the range 1.6 mT÷3.2 mT. The experimental conditions were: $P_{RF}$=400 kW, $P_s$=0.4 Pa, $B_{PG}$=2.4 mT, 80 A ISBI current, 80 A ISBP current, $r_{Cs}$=12 mg/h; plasma pulses lasted 27 s, regularly repeated every 4 min. In the considered range, the negative ion density decreased by increasing the magnetic filter field, in agreement with what found in Cs-free conditions [12]. The accelerated beam current density, also shown in Figure 9, follows the same trend with the magnetic filter field and roughly confirms the basic relationship between negative ion density at the PG and available accelerated current density, as in Figure 8a. Regarding cesium, no significant variation of Cs temperature was observed among the three cases.

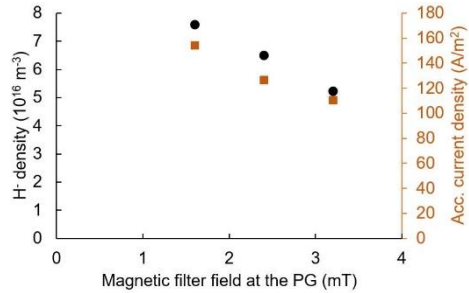

**Figure 9.** H$^-$ density and accelerated current density as a function of the magnetic filter field at the PG. Experimental conditions: $P_{RF}$=400 kW, $P_s$=0.4 Pa (H$_2$), $B_{PG}$=2.4 mT, 80 A ISBI current, 80 A ISBP

A further insight on the effects of the magnetic filter field can be found in the initial phase of each plasma pulse. For a safe ignition of the plasma, RF power ramps up to the desired value in some seconds. $I_{PG}$ is activated before RF power at a minimum value of 0.3 kA, to protect the PG and its Cs layer; $I_{PG}$ increases in the first seconds of the pulse up to the desired value, in ways that were empirically found to guarantee a safe and reliable start for the plasma and the RF plants. Setting $I_{PG}$ at the final value from the beginning would be unfeasible for the plasma ignition. Figure 10a shows $I_{PG}$ and RF1 power (scaled by 1/100 for visualization purposes) as a function of time for one of the plasma pulses in the initial phase of the campaign; the only power of RF1 is plotted as example instead of $P_{RF}$, since the RF generators are not activated at the same time (delays are within 1 s÷2 s). The corresponding evolution of negative ion density with time is shown in Figure 10d; a peaking of negative ion density is observed during the initial increase of $I_{PG}$, indicating that, similarly to Cs free conditions, negative ion density (and presumably plasma density) is peaked below 1.6 mT.

What observed, combined with the results in Cs-free conditions, led to the conclusion that the initial phase of each plasma pulse could be more aggressive towards the Cs layer of the PG surface. The ramp-up of $I_{PG}$ was then modified as in plot b, and finally as in plot c of Figure 10. The negative ion density is shown as a function of time for the two cases in plots e and f. The reduced peaks on H$^-$ density confirm the softer interaction of the plasma with the PG. The remaining peaking of H$^-$ density should not be attributed to variations of gas pressure, which are used in other sources for the onset of the plasma [66], but usually not in SPIDER. Presently, the LAS diagnostic cannot give information on the Cs density within the time scale of these phenomena. For all the three considered plasma pulses, the experimental conditions were: $P_s$=0.4 Pa, 80 A ISBI current, 0 A ISBP current, $r_{Cs}$=7.5 mg/h.

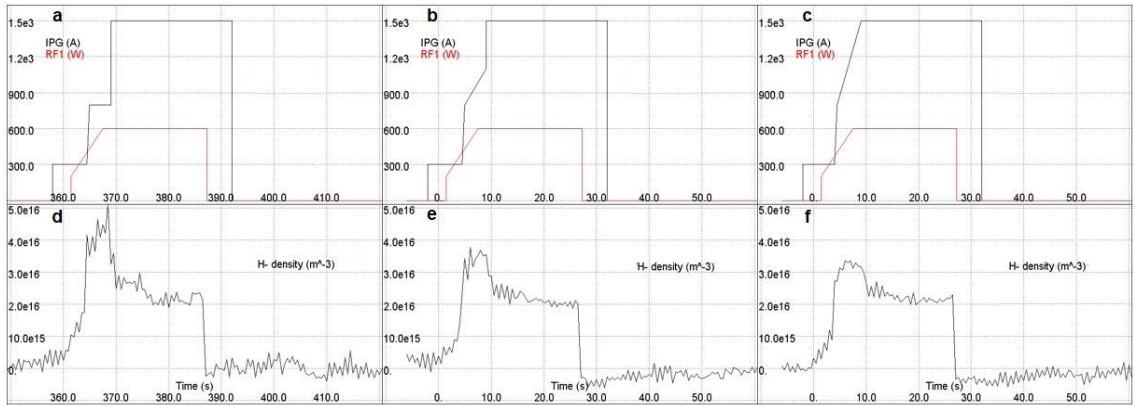

**Figure 10.** Plots a, b and c show the PG current (black) and the RF1 output power scaled by a factor 1/100 (red, all the generators have the same top RF power) for three different plasma pulses, as a function of time. Plots d, e and f show the respective temporal evolution of H⁻ density, as measured by CRDS. Experimental conditions: $P_s$=0.4 Pa (H$_2$), 80 A ISBI current, 0 A ISBP current, $r_{Cs}$=7.5 mg/h.

### 4.4 Influence of BP and PG biasing

In SPIDER, ISBI and ISBP could be independently set in order to reduce the fraction of coextracted electrons; during the experimental campaign it was possible to assess the effectiveness of the two types of biasing. Figure 11a and Figure 11b show the H⁻ density, as measured by CRDS, and the fraction of coextracted electrons, as function of the PG and BP bias currents, i.e. the currents of the ISBI and ISBP power supplies minus the currents flowing on the respective R=0.63 Ω protection resistors (Figure 1a). In the plots of Figure 11, ISBI was set at 0 A, 80 A, 140 A and 190 A, while ISBP was set at 0A, 80 A and 140 A. The corresponding bias currents are lower, due to the plasma interaction with the source elements. At 0 A power supply current, the direction of the corresponding bias current is even reversed. As shown by the plots, the PG-source biasing is much more effective than the BP-source biasing in reducing the fraction of co-extracted electrons. However, the negative ion density is also reduced by the PG-source biasing more than the BP-source one. The best compromise results to keep the PG bias current slightly positive, in the 0 A÷50 A range, similarly to what found in Cs free conditions [12] and in agreement with the studies on the BATMAN test facility, which suggest to apply a bias which is close to the local plasma potential [74]. In the considered measurements, the experimental conditions were: $P_{RF}$=180 kW, $P_s$=0.4 Pa, $B_{PG}$=1.7 mT, $r_{Cs}$=12 mg/h, $U_{ex}$=5 kV, $U_{acc}$=41 kV; plasma pulses lasted 27 s and were repeated after 4 min.. Regarding cesium, the LoS averages Cs density ranged from $1.0 \cdot 10^{14}$ m⁻³ to $1.7 \cdot 10^{14}$ m⁻³, and Cs temperature between 0.27 eV and 0.44 eV, but without clear correlation with respect to the bias currents. Differences may be due to the fact that the plasma pulses were not performed consecutively, but during the experimental sessions of two days.

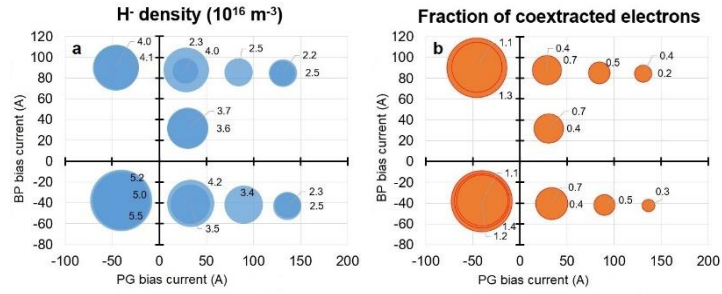

**Figure 11.** H⁻ density (a) and fraction of coextracted electrons (b), as function of PG and BP bias currents (see text). Experimental conditions: $P_{RF}$=180 kW, $P_s$=0.4 Pa (H$_2$), $B_{PG}$=1.7 mT, $r_{Cs}$=12mg/h, $U_{ex}$=5 kV, $U_{acc}$=41 kV.

### 4.5 Operation in deuterium

In SPIDER it was possible for a limited time to study the plasma operation in deuterium. It is known from the previous experimentation in other sources that deuterium requires a higher Cs evaporation rate to keep the same hydrogen performances (or at least get closer to them) in terms of beam current and fraction of co-extracted electrons [[31],[45],[49],[75]-[77]]. R$_{Cs}$ was set to 64 mg/h (in hydrogen, r$_{Cs}$ did not exceed about 25 mg/h). Figure 12 shows, as example, the Cs density measurements from LAS2 as a function of time, for part of the 9$^{th}$ July 2021 experimental session in deuterium. Each peak in plasma density is in correspondence of a plasma pulse; they were performed with $P_{RF}$=400 kW, $P_s$=0.45 Pa, $B_{PG}$=3.2 mT, 190 A ISBI current and 80 A ISBP current. Cs evaporation started at 11:27; unlike what previously observed in hydrogen at $r_{Cs}$≤25 mg/h, the evolution of Cs density is remarkably affected by the sequence of plasma pulses. During each vacuum phase the Cs density grows, but after each plasma pulse the density is reduced. This can be explained by the fact that deuterium atoms, having a higher mass, lead to a higher Cs sputtering yield on the PG-BP, redistributing Cs away from them; in general, the increased sputtering might have involved all surfaces, thus influencing the sticking coefficient of the source walls (which has a clear role as discussed previously in figure 4b). The effects of this phenomenon may have been amplified by the higher Cs evaporation rate. Within the limited time available (two experimental days) to stabilize the Cs conditioning in deuterium, the D⁻ density was not significantly different from the measurements in hydrogen. In Cs-free conditions a significant difference was measured at $B_{PG}$ values close to 1.4 mT÷1.6 mT [[11],[12]]; during the deuterium operation, however, $B_{PG}$ was never below 3.2 mT (as target/flattop value for the plasma pulses) to limit the coextraction of electrons (higher than in hydrogen).

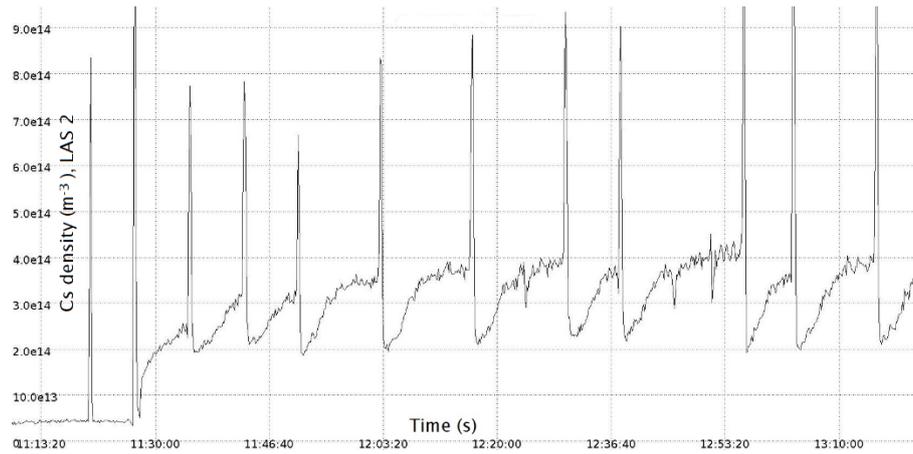

**Figure 12.** Cs density as measured by LoS 2 of the LAS diagnostic, as a function of time, during part of deuterium operation on 9th July 2021. The total Cs evaporation rate was 64 mg/h. Plasma pulses were performed at $P_{RF}$=400 kW, $P_s$=0.45 Pa, $B_{PG}$=3.2 mT, 190 A ISBI current and 80 A ISBP current.

## 5. Discussion of results

The evaporation of Cs in negative ion sources, and the redistribution of Cs caused by the plasma action on the surfaces, are complex but fundamental elements to reach the ITER target values for the beam current (intensity, but also uniformity) and the fraction of coextracted electrons. The LAS and CRDS diagnostics have provided useful information on the spatial distribution of cesium, on the effects of the plasma on the cesiated surfaces and on the generation of negative ions at the PG.

The LAS data showed that the maximum value of average Cs density has a roughly linear dependence on the total Cs evaporation rate from the ovens; nevertheless, the average value of Cs density can strongly vary upon the status of the source and of the cesiation process. The reliability of the source operation will have to be improved by identifying and stabilizing the best operating temperatures in the source, besides improving the vacuum purity against Cs pollutants.

The numerical simulations with AVOCADO, indicating a minor flux of Cs particles at the top and bottom of the source, were basically confirmed by LAS measurements in vacuum phases. The LAS diagnostic showed a further lack of Cs at the bottom, fact that may be explained by the indirect effects on surfaces caused by plasma drifts. This non uniformity could in principle be compensated by setting different evaporation rates among the three Cs oven. However, even if a uniform distribution were obtained in vacuum, plasma would act in a non uniform way in the redistribution, and the latter effect will be dominant for long pulses. In future, possible compensation techniques for Cs non uniformities should also mutually take into account all the other actions that aim at compensating the consequences of plasma drifts on the uniformity of beam current [78]. For example, varying the input RF power among different rows of drivers could also have side effects on the Cs distribution. Changing wall temperatures is known to affect the Cs distribution as well [64].

LAS data showed also that, during plasma pulses, the Cs erosion processes on the surfaces dominate over ionization (and excitation) of cesium, and Cs density is basically linear with RF power (and then also with plasma density). The effects of Cs erosion are even more evident in deuterium, and will have to be taken into account for long plasma pulses (up to 3600 s). The target to maintain the Cs layer stable over long pulses is recognized to be one of the most challenging tasks to reach the ITER targets, and is object of experimentation and numerical simulations [[30],[66],[70],[76]]. In this regard, so called Cs showers on board of the PG are being studied in other research institutes as alternative evaporation systems to guarantee Cs replenishment on the PG and reduce the influence of plasma non-uniformities [[76],[79]].

The study of the Doppler effect on LAS absorption spectra showed that neutral, ground state Cs is not realistically thermalized with hydrogen or deuterium but it's at an higher temperature, of few tenths of electronvolt. Recombination of $Cs^+$ at the BP walls may be the root of the phenomenon; further investigation will be made in the future.

The CRDS diagnostic data have quantified the increment of negative ion production due to the presence of Cs. While surface reactions now dominate over volume reactions, they still basically depend on plasma density, as the dependencies of $H^-$ density on RF power and $B_{PG}$ suggest. As consequence, the basic indications on the operation of the source in Cs free conditions [12] are essentially valid also with cesium. The availability of negative ion density measurements has also helped to better design the $B_{PG}$ pulse profiles for the plasma pulses, in order to better preserve the conditions of the PG surface. At the same time, this kind of transient phenomena further reinforces the need to extend the duration of plasma phases closer and closer to the ITER NBI target of 3600 s.

In Cs-free conditions, the study of PG and BP biasing had led to the conclusion that the sum of PG and BP bias currents have to be slightly positive (0 A÷100 A) to maximize the negative ion density at the PG. This is qualitatively in line also with results and theoretical considerations made in other experiments [74]. With Cs evaporation it was cleared that the PG-source bias dominates over the BP-source biasing in terms of reduction of coextracted electrons, but also that it has a strong effect on negative ions within the explored range of bias voltages. Slightly positive values (0 A÷50 A) of PG bias current can be considered a valid compromise between reducing electrons and not affecting the negative ion density too much (but the optimal values also depend on $B_{PG}$). At last, to get closer to the typical operative conditions of an ITER NBI negative ion source, all the PG apertures will have to be left open during operation, after the improvements of the vacuum pumping system [80][81]. This change may affect the conditions of the plasma in the vicinity of the PG, and the conditions of the PG itself, mainly with an expected reduction of the plasma electronegativity and of the available negative ion current density [69]: however, the role of source parameters, and the main physical dependencies obtained in this investigation will still be valid, permitting a more effective operation of the ITER prototype source SPIDER.

## Acknowledgments

This work has been carried out within the framework of the ITER-RFX Neutral Beam Testing Facility (NBTF) Agreement and has received funding from the ITER Organization. The views and opinions expressed herein do not necessarily reflect those of the ITER Organization.
This work has also been carried out within the framework of the EUROfusion Consortium, partially funded by the European Union via the Euratom Research and Training Programme (Grant Agreement No 101052200 — EUROfusion). The Swiss contribution to this work has been funded by the Swiss State Secretariat for Education, Research and Innovation (SERI). Views and opinions expressed are however those of the author(s) only and do not necessarily reflect those of the European Union, the European Commission or SERI. Neither the European Union nor the European Commission nor SERI can be held responsible for them.

# References

[1] R. S. Hemsworth et al., *Overview of the design of the ITER heating neutral beam injectors*, New J. Phys. 19 025005 (2017).

[2] P. Sonato et al., *The ITER full size plasma source device design*, Fus. Eng. Des. 84 269–274, (2009).

[3] G. Serianni et al., *SPIDER in the roadmap of the ITER neutral beams*, Fus. Eng. Des. 146 2539–2546 (2019).

[4] G. Serianni et al., *First operation in SPIDER and the path to complete MITICA*, Rev. Sci. Instrum. 91, 023510 (2020).

[5] M. Bacal et al., *Negative hydrogen ion production mechanisms*, Appl. Phys. Rev. 2, 021305 (2015).

[6] M. Bacal et al., *Negative ion sources*, J. Appl. Phys. 129, 221101 (2021).

[7] Yu. I. Belchenko et al., *A powerful injector of neutrals with a surface-plasma source of negative ions*, Nucl. Fusion 14, 113 (1974).

[8] Yu. I. Belchenko et al., *Development of Surface-plasma Negative Ions Sources at the Budker Institute of Nuclear Physics*, AIP Conf. Proc. 2052, 030006 (2018).

[9] V. Toigo et al., *On the road to ITER NBIs: SPIDER improvement after first operation and MITICA construction progress*, Fus. Eng. Des. 168, 112622 (2021).

[10] M. Barbisan et al., *Development and first operation of a cavity ring down spectroscopy diagnostic in the negative ion source SPIDER*, Rev. Sci. Instrum. 92, 053507 (2021).

[11] M. Barbisan et al., *Characterization of Cs-free negative ion production in the ion source SPIDER by cavity ring-down spectroscopy*, J. Instrum. 17, C04017 (2022).

[12] M. Barbisan et al., *Negative ion density in the ion source SPIDER in Cs free conditions*, Plasma Phys. Control. Fusion 64, 065004 (2022).

[13] E. Sartori et al., *First operations with caesium of the negative ion source SPIDER*, Nucl. Fusion 62, 086022 (2022).

[14] G. Serianni et al., *Spatially resolved diagnostics for optimization of large ion beam sources*, Rev. Sci. Instrum. 93, 081101 (2022).

[15] O'Keefe et al., Cavity ring-down optical spectrometer for absorption measurements using pulsed laser sources, Rev. Sci. Instrum. 59, 2544 (1988).

[16] R. Pasqualotto et al., *Design of a cavity ring-down spectroscopy diagnostic for negative ion rf source SPIDER*, Rev. Sci. Instrum. 81, 10D710 (2010).

[17] U. Fantz et al., *Optimizing the laser absorption technique for quantification of caesium densities in negative hydrogen ion sources*, J. Phys. D: Appl. Phys. 44, 335202 (2011).

[18] M. Barbisan et al., *Design and preliminary operation of a laser absorption diagnostic for the SPIDER RF source*, Fus. Eng. Des. 146B, pp. 2707-2711 (2019).

[19] M. Barbisan et al., *Laser absorption spectroscopy studies to characterize Cs oven performances for the negative ion source SPIDER*, J. Instrum. 14, C1201 (2019).

[20] K. Tsumori et al., *A review of diagnostic techniques for high-intensity negative ion sources*, Appl. Phys. Rev. 8, 021314 (2021).

[21] M Berger et al., *Cavity ring-down spectroscopy on a high power rf driven source for negative hydrogen ions*, Plasma Sources Sci. Technol. 18 025004 (2009).

[22] A. Mimo et al., *Cavity ring-down spectroscopy system for the evaluation of negative hydrogen ion density at the ELISE test facility*, Rev. Sci. Instrum. 91, 013510 (2020).

[23] C. Wimmer et al., *Beamlet scraping and its influence on the beam divergence at the BATMAN Upgrade test facility*, Rev. Sci. Instrum. 91, 013509 (2020).

[24] F. Grangeon et al., *Applications of the cavity ring-down technique to a large-area rf-plasma reactor*, Plasma Sources Sci. Technol. 8, 448 (1999).

[25] H. Nakano et al., *Cavity Ring-Down System for Density Measurement of Negative Hydrogen Ion on Negative Ion Source*, AIP Conf. Proc. 1390, 359 (2011).

[26] H. Nakano et al., *Cavity Ringdown Technique for negative-hydrogen-ion measurement in ion source for neutral beam injector*, J. Instrum. 11 C03018 (2016).

[27] D. Mukhopadhyay et al., *Quantification of atomic hydrogen anion density in a permanent magnet based helicon ion source (HELEN) by using pulsed ring down spectroscopy*, Rev. Sci. Instrum. 90, 083103 (2019).

[28] R. Agnello et al., *Cavity ring-down spectroscopy to measure negative ion density in a helicon plasma source for fusion neutral beams*, Rev. Sci. Instrum. 89, 103504 (2018).

[29] B. Heinemann et al., *Towards large and powerful radio frequency driven negative ion sources for fusion*, New J. Phys. 19, 015001 (2017).

[30] A. Mimo et al., *Studies of Cs dynamics in large ion sources using the CsFlow3D code*, AIP Conf. Proc. 2052, 040009 (2018).

[31] D. Wünderlich et al., *Formation of large negative deuterium ion beams at ELISE*, Rev. Sci. Instrum. 90, 113304 (2019).

[32] S. Cristofaro et al., *Correlation of Cs flux and work function of a converter surface during long plasma exposure for negative ion sources in view of ITER*, Plasma Res. Express 2, 035009 (2020).

[33] M. Bigi et al., *Design, manufacture and factory testing of the Ion Source and Extraction Power Supplies for the SPIDER experiment*, Fus. Eng. Des. 96-97, 405-410 (2015).

[34] A. Rizzolo et al., *Characterization of the SPIDER Cs oven prototype in the CAesium Test Stand for the ITER HNB negative ion sources*, Fus. Eng. Des. 146A, pp. 676-679 (2019).

[35] S. Cristofaro et al., *Design and comparison of the Cs ovens for the test facilities ELISE and SPIDER*, Rev. Sci. Instrum. 90, 113504 (2019).

[36] M. De Muri et al., *SPIDER Cs Ovens functional tests*, Fus. Eng. Des. 161, 112331 (2021).

[37] E Sartori, *Simulation-based quantification of alkali-metal evaporation rate and systematic errors from current–voltage characteristics of Langmuir–Taylor detectors*, IEEE Transactions on Instrumentation and Measurement 69 (7), 4975-4986 (2020).

[38] N. Marconato et al., *An optimized and flexible configuration for the magnetic filter in the SPIDER experiment*, Fus. Eng. Des. 166, 112281 (2021).

[39] U. Fantz et al., *Spectroscopy—a powerful diagnostic tool in source development*, Nucl. Fusion 46 S297–S306 (2006).

[40] P. McNeely et al., *A Langmuir probe system for high power RF-driven negative ion sources on high potential*, Plasma Sources Sci. Technol. 18 014011 (2009).

[41] U. Fantz et al., *A comparison of hydrogen and deuterium plasmas in the IPP prototype ion source for fusion*, AIP Conf. Proc. 1515, 187 (2013).

[42] U. Fantz et al., *Physical performance analysis and progress of the development of the negative ion RF source for the ITER NBI system*, Nucl. Fusion 49 125007 (2009).

[43] K. W. Ehlers and K. N. Leung, *Effect of a magnetic filter on hydrogen ion species in a multicusp ion source*, Rev. Sci. Instrum. 52, 1452 (1981).

[44] R. K. Janev et al., Elementary processes in hydrogen-helium plasmas: Cross sections and reaction rate coefficients, Springer-Verlag Berlin Heidelberg (1987).

[45] E. Speth et al., *Overview of the RF source development programme at IPP Garching*, Nucl. Fusion 46 S220–S238 (2006).

[46] P. Franzen, *Progress of the ELISE test facility: results of caesium operation with low RF power*, Nucl. Fusion 55, 053005 (2015).

[47] Yu. Belchenko et al., *Effect of plasma grid bias on extracted currents in the RF driven surface-plasma negative ion source*, Rev. Sci. Instrum. 87, 02B119 (2016).

[48] K. Tsumori et al., *Negative ion production and beam extraction processes in a large ion source*, Rev. Sci. Instrum. 87, 02B936 (2016).

[49] K. Ikeda et al., *Exploring deuterium beam operation and the behavior of the co-extracted electron current in a negative-ion-based neutral beam injector*, Nucl. Fusion 59 076009 (2019).

[50] G. Fubiani et al., *Modeling of plasma transport and negative ion extraction in a magnetized radio-frequency plasma source*, New J. Phys. 19 015002 (2017).

[51] M. Pavei et al., *SPIDER plasma grid masking for reducing gas conductance and pressure in the vacuum vessel*, Fus. Eng. Des. 161, 112036 (2020).

[52] P. Agostinetti et al., *Physics and engineering design of the accelerator and electron dump for SPIDER*, Nucl. Fusion 51 063004 (2011).

[53] G. Serianni et al., *Numerical simulations of the first operational conditions of the negative ion test facility SPIDER*, Rev. Sci. Instrum. 87, 02B927 (2016).

[54] C.F. Barnett, J.A. Ray, E. Ricci, M.I. Wilker, E.W. McDaniel, E.W. Thomas, H. B. Gilbody, *Atomic data for controlled fusion research*, technical report ORNL-5206(Vol.1), Oak Ridge National Laboratories (1977).

[55] C. Poggi et al., *Langmuir probes as a tool to investigate plasma uniformity in a large negative ion source*, IEEE Transactions on Plasma Science 50, n. 11, pp. 3890-3896 (2022).

[56] L. Schiesko et al., *Magnetic field dependence of the plasma properties in a negative hydrogen ion source for fusion*, Plasma Phys. Control. Fusion 54 105002 (2012).

[57] S. Lishev et al., *Spatial distribution of the plasma parameters in the RF negative ion source prototype for fusion*, AIP Conf. Proc. 1655, 040010 (2015).

[58] S. Lishev et al., *Influence of the configuration of the magnetic filter field on the discharge structure in the RF driven negative ion source prototype for fusion*, AIP Conf. Proc. 1869, 030042 (2017).

[59] V Candeloro, et al., *Electron scraping and electron temperature reduction by bias electrode at the extraction region of a large negative ion source*, IEEE Transactions on Plasma Science 50, n. 11, pp. 3983-3988 (2022).

[60] N. Marconato et al., *Numerical and experimental assessment of the new magnetic field configuration in SPIDER*, IEEE Transactions on Plasma Science 50, n. 11, pp. 3884-3889 (2022)..

[61] W. Demtröder, *Laser Spectroscopy 1: Basic Principles*, Springer (2014).

[62] G. Serianni et al., *SPIDER, the Negative Ion Source Prototype for ITER: Overview of Operations and Cesium Injection*, IEEE Transactions on Plasma Science 51, n. 3, pp. 927-935 (2023).

[63] A. T. Forrester, *Large ion beams: fundamentals of generation and propagation*, Wiley-Interscience New York (1988).

[64] R. Gutser et al., *Dynamics of the transport of ionic and atomic cesium in radio frequency-driven ion sources for ITER neutral beam injection*, Plasma Phys. Control. Fusion 53 105014 (2011).

[65] A. Mimo et al., *Modelling of caesium dynamics in the negative ion sources at BATMAN and ELISE*, AIP Conf. Proc. 1869, 030019 (2017).

[66] C. Wimmer et al., *Improved understanding of the Cs dynamics in large H− sources by combining TDLAS measurements and modeling*, AIP Conf. Proc. 2011, 060001 (2018).

[67] E. Sartori et al., *Avocado: A numerical code to calculate gas pressure distribution*, Vacuum 90, pp. 80–88, (2013).

[68] M. Fadone et al., *Interpreting the dynamic equilibrium during evaporation in a cesium environment*, Rev. Sci. Instrum. 91, 013332 (2020).

[69] E. Sartori et al., *Influence of plasma grid-masking on the results of early SPIDER operation*, Fus. Eng. Des. 194, 113730 (2023).

[70] D. Wünderlich et al., *Long pulse operation at ELISE: Approaching the ITER parameters*, AIP Conf. Proc. 2052, 040001 (2018).

[71] B. P. Duteil, *Development of a Collisional Radiative Model for Hydrogen-Cesium Plasmas and Its Application to SPIDER*, IEEE Transactions on Plasma Science 50, n. 11, pp. 3995-4001 (2022).

[72] B. Zaniol et al., *First measurements of optical emission spectroscopy on SPIDER negative ion source*, Rev. Sci. Instrum. 91, 013103 (2020).

[73] E. Sartori et al., Development of a set of movable electrostatic probes to characterize the plasma in the ITER neutral beam negative-ion source prototype, Fus. Eng. Des. 169, 112424 (2021).

[74] C. Wimmer, *Extraction of negative charges from an ion source: Transition from an electron repelling to an electron attracting plasma close to the extraction surface*, Journal of Applied Physics 120, 073301 (2016).

[75] D. Wünderlich et al., *Initial caesium conditioning in deuterium of the ELISE negative ion source*, Plasma Phys. Control. Fusion 60 085007 (2018).

[76] D. Wünderlich et al., *NNBI for ITER: status of long pulses in deuterium at the test facilities BATMAN Upgrade and ELISE*, Nucl. Fusion 61 096023 (2021).

[77] M. Bacal et al., *Negative ion source operation with deuterium*, Plasma Sources Sci. Technol. 29 033001 (2020).

[78] E. Sartori et al., *Experimental results of the SPIDER negative ion accelerator in view of the next operations*, 8th International symposium on Negative Ions, Beams and Sources - NIBS'22 (2-7 October 2022). Contributed paper submitted to Journal of Instrumentation.

[79] S. G. Kostantinov, *A System of Distributed Cesium Feeding for Increasing the Efficiency of Powerful Sources of Negative Hydrogen Ions*, Instruments and Experimental Techniques 60, pp. 74–77 (2017).

[80] D. Marcuzzi et al., *Lessons learned after three years of SPIDER operation and the first MITICA integrated tests*, Fus. Eng. Des. 191, 113590 (2023).

[81] E Sartori, M Siragusa, G Berton, C Cavallini, S Dal Bello, M Fadone, et al., *Design of a large nonevaporable getter pump for the full size ITER beam source prototype*, Journal of Vacuum Science & Technology B 41 (3) (2023)